\documentstyle[preprint,eqsecnum,aps]{revtex}
\tightenlines
\draft
\preprint{hep-ph/9808290}

\catcode`\@=11
\def\@citex[#1]#2{%
\if@filesw \immediate \write \@auxout {\string \citation {#2}}\fi
\@tempcntb\m@ne \let\@h@ld\relax \def\@citea{}%
\@cite{%
  \@for \@citeb:=#2\do {%
    \@ifundefined {b@\@citeb}%
      {\@h@ld\@citea\@tempcntb\m@ne{\bf ?}%
      \@warning {Citation `\@citeb ' on page \thepage \space
undefined}}%
      {\@tempcnta\@tempcntb \advance\@tempcnta\@ne%
      \@tempcntb\number\csname b@\@citeb \endcsname \relax%
      \ifnum\@tempcnta=\@tempcntb 
	\ifx\@h@ld\relax%
	  \edef \@h@ld{\@citea\csname b@\@citeb\endcsname}%
	\else%
	  \edef\@h@ld{\ifmmode{-}\else--\fi\csname
b@\@citeb\endcsname}%
	\fi%
      \else
	\@h@ld\@citea\csname b@\@citeb \endcsname%
	\let\@h@ld\relax%
      \fi}%
    \def\@citea{,\penalty\@highpenalty\,}%
  }\@h@ld
}{#1}}

\def\@citeb#1#2{{[#1]\if@tempswa , #2\fi}}
\def\@citeu#1#2{{$^{#1}$\if@tempswa , #2\fi }}
\def\@citep#1#2{{#1\if@tempswa , #2\fi}}

\def\bcites{         
	\catcode`\@=11
	\let\@cite=\@citeb
	\catcode`\@=12
}

\def\upcites{         
	\catcode`\@=11
	\let\@cite=\@citeu
	\catcode`\@=12
}

\def\plaincites{      
	\catcode`\@=11
	\let\@cite=\@citep
	\catcode`\@=12
}

\newcommand{\cP}{{\cal P}}
\newcommand{\CovD}{D}

\newcommand{\Fs}[1]{\hat{{#1}}}
\newcommand{\fs}[1]{\hat{{#1}}}

\newcommand{\cQ}{{\cal Q}}
\newcommand{\Klein}{\cP^2+m_c^2}
\newcommand{\Tr}{{\rm Tr}}
\newcommand{\tr}{{\rm tr}}
\newcommand{\bra}[1]{\langle{#1}|}
\newcommand{\ket}[1]{|{#1}\rangle}
\newcommand{\braket}[2]{\langle{#1}|{#2}\rangle}
\newcommand{\sG}{\sigma G}

\newcommand{\be}{\begin{equation}}
\newcommand{\ee}{\end{equation}}
\newcommand{\bea}{\begin{eqnarray}}
\newcommand{\eea}{\end{eqnarray}}
\newcommand{\ba}{\begin{array}{l}}
\newcommand{\ea}{\end{array}}

\newcommand{\bb}{\begin{thebibliography}}
\newcommand{\eb}{\end{thebibliography}}
\newcommand{\ci}[1]{\cite{#1}}
\newcommand{\lab}[1]{\label{#1}}
\newcommand{\re}[1]{(\ref{#1})}
\newcommand{\Ds}{\displaystyle}
\newcommand{\half}{{\textstyle{\frac{1}{2}}}}

\newcommand{\Dirac}{\Fs{\cP}+im_c}

\begin{document}


\title{Axial Currents of Virtual Charm   in Light Quark Processes}
\author{F. ~Araki, M. ~Musakhanov\footnote
{Associate Member of ICTP,
permanent address: Theoretical  Physics Dept, Tashkent State University,
 Tashkent 700095, Uzbekistan} and H. ~Toki}
\address{Research Center for Nuclear Physics (RCNP), Osaka  University,
Ibaraki, Osaka 567-0047, Japan}
\date{\today}
\maketitle
\begin{abstract}
The systematic investigations of the role of 
the virtual charm axial currents in the decay of $B$-mesons to
$\eta '$ $K$ and the spin structure of the nucleon are performed. 
We reduce the divergence of the virtual charm axial current
to a specific gluon operator. 
Since this operator receives the main contribution from topologically
nontrivial components of the QCD vacuum, we rederive the Diakonov-Petrov
Effective Action, based on the instanton QCD vacuum model. This action is 
applied to the calculations of the coupling of  $\eta '$  
to the charm axial current and found
$f_{\eta '}^{(c)} (\mu\,\simeq m_c)\,=\,- (12.3\,\sim 18.4\, {\rm MeV}), $ 
providing       
a possibility for the explanations of the recent experimental data on
$B\,\to\,\eta ' \, K$-decay. Analogous calculations of the virtual charm 
content of the nucleon spin leads to $\Delta c  (\mu\,\simeq m_c)
\,=\,-\, (0.015\,\sim\,0.024),$ 
which is one order of magnitude smaller than the 
analogous contribution of the strange quark.
\end{abstract}


\narrowtext
\section{Introduction}

The processes sensitive to the OZI rule violations and the contributions of 
the nonvalence components of hadrons become of great interest.
In the present paper, we would like to discuss 
the role of the charm content of 
the light hadrons in the $B$-meson decay to $\eta '$ and $K$-mesons 
and in deep--inelastic scattering with polarized charged leptons 
on polarized nucleons. 
In  both cases there are no charmed hadrons both in the initial 
and the final states, which
means that a charm may give a contribution only through virtual processes.
\\ \\
\underline { Decay of $B$-mesons to
$\eta '$ $K$}

Recently there is a great theoretical interest 
\ci{HZ,SZ,AG97,ACGK97,FK97,FK98,F,CT,CT98,DKY,P} 
on the new
experimental data on the branching ratios of the 
decays of $B\,\rightarrow \,\eta ' \, K $\ci{CLEO1,CLEO2}: 
\begin{eqnarray}
\label{etapkpm}
Br( B^\pm \rightarrow \eta^{\prime}  K^\pm) & = & (6.5_{- 1.4}^{+ 1.5}
\pm  0.9) \times 10^{-5},
\\
Br(B^0 \to \eta^{\prime} K^0 ) & = & (4.7_{-2.0}^{+2.7} \pm 0.9) \times 
10^{-5}~. 
\end{eqnarray}
In the Standard Model, the Cabibbo favored $b\,\rightarrow \,\bar c c s$
elementary process may be followed by conversion of $\bar c c$ pair into
$\eta '$ through  gluons.
The amplitude of this process is described by  
\be
\label{amplitude}
M = \frac{G_{F}}{\sqrt{2}} V_{cb} V_{cs}^* a_1 
\bra{\eta'(p)}\bar c\gamma_{\mu}\gamma_{5}c \ket{0}
\bra{K(q)}\bar s \gamma_{\mu} b \ket{B(p+q)} .
\ee
Here $G_F$ is the weak coupling constant,  $V_{cb}$, $ V_{cs}^*$ are
Kobayashi-Masukawa matrix elements and $a_{1}=0.25$ is the 
phenomenological number
 obtained from a fit to experiments (see \ci{HZ} for the references). 
The matrix element 
\be
\bra{\eta ' (p)}\bar c\gamma_{\mu}\gamma_{5}c\ket{0}  =
 -i f_{\eta '}^{(c)} p_{\mu}
\lab{fc} 
\ee
is non-zero due to the virtual $ \bar c c \rightarrow  gluons $ transitions. 
Certainly, this matrix element is suppressed by  $ 1/m_{c}^2 $ factor. 
However,
due to the presence of strong nonperturbative  gluon fields in the QCD vacuum 
 together with
 the Cabibbo favored  $ b \rightarrow c $ transition 
the suggested $b\,\rightarrow \,\bar c c s$ mechanism \re{amplitude} may be
expected to  compete appreciably 
with the other mechanisms  of the 
$ B \rightarrow K \eta' $ process \ci{F,DKY}.

If we assume the dominance of the virtual charm mechanism \re{amplitude},  
   the branching ratio is written in terms of 
 $f_{\eta'}^{(c)}$ as \ci{HZ}
 \be
\label{c}
Br( B \rightarrow K \eta') \simeq 3.92 \cdot 10^{-3} \cdot
\left( \frac{f_{\eta'}^{(c)}}{1 \; {\rm GeV}} \right)^2 .
\ee
 Using the experimental data, \re{etapkpm}, it is found 
$f_{\eta'}^{(c)} \simeq 140 \; {\rm MeV} \; \; (``exp").$ 

This value perfectly coincides with the estimate
of Halperin and Zhitnitsky \ci{HZ}:
\be
\label{HZestimation}
f_{\eta'}^{(c)}  = ( 50 - 180) \; {\rm MeV}.
\ee
On the other hand,  a recent phenomenological study placed another
bound on $f_{\eta^\prime}^{(c)}$, namely $-65 ~\mbox{{\rm MeV}} \leq
f_{\eta^\prime}^{(c)} \leq 15  ~\mbox{{\rm MeV}}$, 
with $f_{\eta '}^{(c)}$ being consistent with zero by analyzing
the $Q^2$ evolution of the  $\eta^\prime \gamma$ form
factor \cite{FK97}, and more recently it was estimated from observed 
ratio of $J/\psi$ decay to $\eta '$ and $\eta_c$ the value of
$f_{\eta '}^{(c)}= -(6.3 \pm 0.6){\rm MeV}$ \cite{FK98}. 
Other similar estimation 
which leads to $|f_{\eta'}^{(c)} | < 12 {\rm MeV}$  was
made in \ci{P}. 
Ali et al.  considered the complete
amplitude for the exclusive $B-$meson decays, including 
 $\eta^{\prime}  K$ channels, 
where they combined the contribution from the 
process $b \to s (c\bar{c}) \to s(gluons) \to 
s \eta^{(')}$  with all the others 
\cite{AG97,ACGK97}.
Their estimations    gave
 $|f_{\eta^\prime}^{(c)}| \simeq 5.8$ {\rm MeV}\ci{AG97}
and $f_{\eta^\prime}^{(c)} = -3.1$ ($-2.3$) {\rm MeV} (for 
$m_c$ in the range $1.3$ - $1.5$ {\rm GeV})\ci{ACGK97}  in agreement with the
analysis\cite{FK97}. They stressed  the importance of the 
sign of 
$f_{\eta^\prime}^{(c)}$ and found a theoretical branching ratio
in the range
$$Br( B \rightarrow \eta^{\prime}  K) \,=\,(2-4) \times 10^{-5},$$
which is somewhat smaller than the experimental one \re{etapkpm}.
The similar analysis made in \ci{CT} led the authors to conclude that 
$f_{\eta^\prime}^{(c)}= - 50 {\rm MeV}$ 
might provide the explanation of the data.
 
Having this situation, it is important to recalculate $f_{\eta^\prime}^{(c)}$ 
to clarify the 
mechanism of $ B \rightarrow K \eta' $ decay 
in the similar framework performed by Halperin and Zhitnitsky \ci{HZ}.
\\ \\
\underline{Polarized DIS}

In the constituent quark model, the  spin of the proton is supposed  to be
carried by
 $q$ = $u, d$ valence quarks so that $\Delta\Sigma= \Delta u + \Delta d = 1$.
The quark spin $\Delta q$ is defined as  
\be
\Delta q\,2m_{N}\,s^\mu = \langle p,s \vert \overline q \gamma^\mu 
\gamma_5 q \vert p,s\rangle ,
\lab{def1}
\ee
where $m_{N}$ and $s^\mu$ are the mass and the spin
of the nucleon, respectively. On the other hand, 
$$\Delta q \,=\, \int^1_0 dx\, \left(q_R(x)-q_L(x)\right),$$  
where
$q_{R(L)}(x)$ are quark distributions of chirally right-handed 
(left-handed) quarks in a polarized proton.  

Deep inelastic scattering (DIS) with polarized
charged leptons on polarized targets provides the investigations of
 the quark distributions
$q_{R(L)}$.  These quantities are extracted from 
the structure function $g_1(x,Q^2)$ measured in polarized DIS 
using the parton model relation
$g_1(x,Q^2) =\frac{1}{2}\sum_q  e_q^2\,\left(q_R(x)-q_L(x)\right)$.

So called "spin crisis" problem is related with a very large disagreement 
between the experiments 
and the prediction of the naive constituent quark model 
for the first 
moment of the proton(neutron) spin structure function $\Gamma_{1}^{p(n)}$
defined by 
\begin{equation}
\Gamma_{1}^{p(n)} \equiv \int dx g_{1}^{p(n)}(x) 
= +(-) \frac{1}{12} g_{A}^{3} + \frac{1}{36} g_{A}^{8}
+ \frac{1}{9} g_{A}^{0}. 
\label{naiveg1}
\end{equation}
Here, $$g_{A}^{3} =\Delta u - \Delta d = 1.25$$  
and $$g_{A}^{8}= (\Delta u + \Delta d -2 \Delta s)= 0.69$$ are 
the isovector and the octet axial charges  measured from neutron and hyperon
decays and  $$g_{A}^{0}= \Delta \Sigma = (\Delta u + \Delta d + \Delta s).$$
The expectation was $g_{A}^{0} = 1$ but the 
measured value of this quantity given by EMC in 1988 implied
that $g_{A}^{0} \sim  0.12$. It means that
only $\sim 12 \%$ of the spin of the proton is carried by its quarks!
The modern value is $g_{A}^{0}\sim  0.3$\cite{EK95,Mulders98,Ioffe98}.

In all of the consideration of the nucleon spin the contribution of the charm 
was neglected completely. With the account of
this contribution
  $$g_{A}^{0}=\Delta \Sigma + 2\Delta c$$ 
  and  we have now the problem to 
estimate also $\Delta c.$ We may expect sizable value of $\Delta c,$
since in this case we are
dealing again with axial currents which may be strongly affected by the vacuum
 nonperturbative  gluon fields. 
 The previous calculations of $\Delta c$
by Halperin and Zhitnitsky \ci{HZ97-dis}
and Blotz and Shuryak \ci{BS97}
gave quit different results
$\Delta c\sim 0.3$ \ci{HZ97-dis} and 
$\Delta c/\Delta \Sigma \, =\, -\,(0.2\sim 0.08)$\ci{BS97}. They contradicted
each other in the sign and also absolute value. This is
our motivation for the recalculation of the charm contribution to the
spin of the nucleon.
\\ \\
\underline{Axial currents of virtual charm}

The symmetry of  the classical 
lagrangian may be destroyed by quantum  fluctuations \ci{Schw,Adl,BJa}. 
In gauge theories the axial anomaly arises from noninvariance of 
the fermionic measure against axial transformations 
in the path integrals of the theory\ci{Fujikawa} (see also ref.\ci{FMP},
concerning higher-loop corrections). The present problem is intimately related
with this phenomenon.

 In the following we will work only with the Euclidean QCD, where its
 convention is written in the footnote\footnote
{$
ix_{M0}=x_{E4},\,\,\,x_{Mi}=x_{Ei},\,\,\,A_{M0}=iA_{E4},\,\,\,
A_{Mi}= - A_{Ei},\,\,\, 
 \psi_{M}=\psi_{E},\,\,\, i\bar\psi_{M}=\psi_{E}^{\dagger},\,\,\,
 \gamma_{M0}=\gamma_{E4} \\ \gamma_{Mi}=i\gamma_{Ei},\,\,\,
 \gamma_{M5}=\gamma_{E5}.
$ In the following we will omit index $E.$}.    
In the Euclidean QCD the axial anomaly in the light quark axial
 current in the chiral limit reads
\be
\partial_{\mu}\psi^{\dagger}_{f}\gamma_{5}\gamma_{\mu}\psi_{f} 
= - i  \frac{g^2}{16\pi^2} G\tilde G,
\lab{div}
\ee
where $\psi_{f}$ is the light quark field
$ (f=u,d,s)$ and
 $g$ is the QCD coupling constant.  
 $2G\tilde G= \epsilon^{\mu\nu\lambda\sigma}
 G^{a}_{\mu\nu}  G^{a}_{\lambda\sigma}$, where $G^{a}_{\mu\nu}$ is
the gluon field strength operator with $a$ being the color index.

The situation with heavy quarks is very different, since we must
take into account the contribution of the mass term. 
The divergence of the axial current of
charmed quarks has a form:
\be
\partial_{\mu}c^{\dagger}\gamma_{\mu}\gamma_{5}c 
= - i  \frac{g^2}{16\pi^2} G\tilde G    
+ 2m_{c}c^{\dagger}\gamma_{5}c,
\lab{divc}
\ee
The first term in \re{divc} again comes from noninvariance of 
the fermionic measure (or in other words - from Pauli-Villars regularization).
The main problem here is to calculate the contribution from the 
second term in \re{divc}. 
It is clear that this one is reduced to the problem of the calculation
of the vacuum expectation value of the operator 
$2m_{c}c^{\dagger}\gamma_{5}c$
 in the  presence of a gluon fields.

In the path integral approach, the calculation of 
the contribution of this term to any matrix element over light hadrons 
may be considered in the sequence of the integrations. 
In the first step, the integration over $c$-quark is performed, 
and the next step is
the calculation of the integral over  gluon fields and finally the integration 
over light quarks.

We consider here the first step -- the integration over c-quarks.
The resulting $\langle 2m_{c}c^{\dagger}\gamma_{5}c\rangle$
 must be a gauge invariant function of the gauge field $A$
and therefore must be expressed through  the gluon field strength tensor and
their covariant derivatives.
The matrix elements 
of such types of the  operators are almost completely defined  
by nonperturbative topologically nontrivial (like instantons) 
contributions at least for the small virtualities flowing
through these operators, which is our case. This one can be  clearly
justified by the consideration of low-energy theorems for the 
various matrix elements of the gluonic operator in \re{div} \ci{Shi}.
It was shown that only with account of the instanton-like configurations of 
vacuum gluonic fields, it is possible to satisfy these theorems and that 
Diakonov \& Petrov (DP) chiral quark Effective Action \ci{DP,DPW} based
on the QCD instanton vacuum model \ci{Shu82,DP84}
reproduce well these low-energy theorems in the chiral limit but fail 
beyond this limit \ci{MK,SM}.
To our present knowledge the instanton structure of QCD vacuum is 
characterized by the average size $\rho$ and by the average inter-instanton 
distance $R$, which are \ci{DP84,Shu82}
\be
\rho =1/3\; {\rm fm}, \, \, \, \,\,\,\, R=1\;{\rm fm} .
\label{rho,R}    
\ee
Therefore the packing parameter $(\rho /R)^4 = 0.012$ is small, legitimatizing
independent averaging over positions and orientations of the instantons.   
  
In the present paper  we first calculate gluonic operators 
in the divergency of the virtual charm axial currents, further we 
rederive the DP Effective Action starting from the Lee\&Bardeen result for the
quark propagator in the instanton media and apply this action for the
calculations of the correlators with  gluonic operators.  
Finally we apply these results to the calculations of the virtual charm effects
in above mentioned problems of the $B$-decay and DIS.

\section{The divergence of the virtual charm axial currents} 
 
We calculate the expectation value of the operator 
$2m_c c^\dagger \gamma_5 c$ in the presence of  gluon field.
For this, we take the  method, which  was 
developed by Schwinger in electrodynamics 
many years ago\cite{Schw}, and later  was applied to QCD 
by Vainshtein et al. \cite{VZNS83}.
The key point of this method is based on an assumption of 
a possibility of an expansion of the Green function such as 
$\langle 2m_c c^\dagger \gamma_5 c\rangle$ over $G/m_c^2$.

We introduce first the coordinate and momentum operators,
$X_\mu$ and $p_\mu$, respectively, which satisfy
$[p_\mu, X_\nu]=i\delta_{\mu\nu}\ ,\ [p_\mu,p_\nu]=[X_\mu,X_\nu]=0$.
We define then the covariant momentum operator $\cP_\mu$
satisfying the following commutation relations,
\begin{equation}
[\cP_\mu, X_\nu]=i\delta_{\mu\nu}\;,\;
[\cP_\mu,\cP_\nu]=igG_{\mu\nu}^at^a\ ,\label{eq:op1}
\end{equation}
where $t^a$ is a generator of the color group 
and $G_{\mu\nu}^a$ is the gluon field strength tensor.
Moreover, we introduce a formal complete set of states $\ket{x}$
as the eigenstates of the coordinate operator $X_\mu$,
\begin{equation}
X_\mu\ket{x}=x_\mu\ket{x}\ ,
\end{equation}
which satisfies 
\begin{equation}
\braket{y}{x}=\delta^{(4)}(x-y)\;\;,\;\;
\int d^4 x\ket{x}\bra{x}=1\ .
\end{equation}
In this basis, the operator $\cP_\mu$ acts 
as a covariant derivative $\CovD_\mu$,
\begin{equation}
\bra{y}\cP_\mu\ket{x}=i\CovD_\mu\braket{y}{x}
=\left(i\partial_\mu+gA_\mu^a(x)t^a\right)\delta^{(4)}(x-y)\ .
\end{equation}
The algebra (\ref{eq:op1}) is the basic tool of the Schwinger formalism.
We expand the Green functions in the gluon background field, 
and need to use only this algebra in each order of expansion.

Next, we calculate the expectation value of the operator 
$2m_c c^\dagger(x) \gamma_5 c(x)$,
which is the Green function that should be expanded 
by using the Schwinger method.
We consider the integration over $c$-quarks.
In the path integral approach, we define :
\begin{equation}
\langle 2m_c c^\dagger(x)\gamma_5 c(x)\rangle=
\int { D} c {D}c^\dagger 
2m_c c^\dagger (x)\gamma_5 c(x) \exp \left\{
\int d^4 y \ c^\dagger (y)(\Dirac) c(y)\right\}\label{eq:1}
\end{equation}
with $\Fs{\cP}\equiv\cP_\mu \gamma_\mu$.
Since the argument of the exponential is quadratic in the quark field $c$,
the path integral (\ref{eq:1}) can be written in the form,
\begin{equation}
\langle2m_c c^\dagger (x)\gamma_5 c(x)\rangle
=2m_c \det\|\Dirac\|
\ \bra{x}\Tr \gamma_5\frac{1}{\Dirac}\ket{x}\ ,\label{eq:2}
\end{equation}
where $\Tr$ denotes the trace over spin and color indices.
Since $x$ is a continuous variable, 
the operator has an infinite number of matrix elements,
and in calculating the determinant of this matrix
there arise infinities of various types.
Hence, the determinant, $\det\|\Dirac\|$, must be regularized
in the standard manner as
\begin{displaymath}
\det\|\Dirac\|\longrightarrow
\det\left\|\frac{(\Dirac)(\fs{p}+iM)}{(\fs{p}+im_c)(\Fs{\cP}+iM)}\right\|\ ,
\end{displaymath}
where $M$ is the Pauli-Villars regulator mass.
Eq.(\ref{eq:2}) must be gauge invariant and expressed 
through the gluon field strength tensor and the covariant derivatives.

Eq.(\ref{eq:2}) is expanded in the series of a power of $G/m_c^2$
under the assumption that the field strength $G_{\mu\nu}^a$ is
much less than the square of $c$-quark mass $m_c^2$.
Here, we will take into account ${\cal O}(G^2)$ and ${\cal O}(G^3)$ terms
in the expansion of (\ref{eq:2}).
We start from the calculation of
\begin{equation}
H(x)\equiv
2m_c\bra{x}\Tr\gamma_5\frac{1}{\Dirac}\ket{x}\ .\label{eq:4}
\end{equation}
The calculations of the $\det\|\Dirac\|$ is presented in the Appendix, 
since it give a contribution to \re{eq:2} starting ${\cal O}(G^4)$ terms.

Using the formulas
\begin{displaymath}
\Fs{\cP}\ ^2=\cP^2+\frac{g}{2}\sG\quad
{\rm and}\quad I=\frac{1}{\Fs{\cP}-im_c}(\Fs{\cP}-im_c)\ ,
\end{displaymath}
where $\sG\equiv\sigma_{\mu\nu}G_{\mu\nu}$,
$\sigma_{\mu\nu}\equiv\frac{i}{2}[\gamma_\mu , \gamma_\nu]$
and $I$ is an identity matrix,
Eq.(\ref{eq:4}) reduces to
\begin{eqnarray}
H(x)&=&2m_c\bra{x}\Tr\gamma_5\frac{1}{\Dirac}
\frac{1}{\Fs{\cP}-im_c}(\Fs{\cP}-im_c)\ket{x}\nonumber\\
&=&-2im_c^2\bra{x}\Tr\gamma_5
\frac{1}{\Klein+\frac{g}{2}\sG}\ket{x}\nonumber\\
&=&-2im_c^2\bra{x}\Tr\gamma_5
\left\{
\frac{1}{\Klein}\frac{g}{2}\sG
\frac{1}{\Klein}\frac{g}{2}\sG
\frac{1}{\Klein}\right.\nonumber\\
&&\left.\;\; 
-\ \frac{1}{\Klein}\frac{g}{2}\sG
\frac{1}{\Klein}\frac{g}{2}\sG
\frac{1}{\Klein}\frac{g}{2}\sG
\frac{1}{\Klein}
+\cdots
\right\}\ket{x}\label{eq:6}\\
&\equiv&H_2(x)+H_3(x)+\cdots\ .\nonumber
\end{eqnarray}
It has been used here that the trace of an odd product 
of $\gamma$ matrices vanishes.

It is straightforward to calculate the second term of 
the right hand side of (\ref{eq:6}),
if neglect the noncommutativity of the operators
$\cP_\mu$ and $G_{\mu\nu}$ in (\ref{eq:6}).
Thus,
\begin{displaymath}
H_3(x)
=\frac{ig^3m_c^2}{2^2}\bra{x}\Tr\gamma_5
\frac{1}{(\Klein)^4}(\sG)^3\ket{x}\ .
\end{displaymath}
In that case,
since the operator $\cP_\mu$ can be replaced 
by the ordinary momentum $p_\mu$,
we can use the evident formulas,
\begin{eqnarray}
\bra{x}\frac{1}{(\Klein)^n}\ket{x}
&=&\int\frac{d^4p}{(2\pi)^4}\left.\frac{1}{(\Klein)^n}\right|_{A=0}
\nonumber\\
&=&\int\frac{d^4p}{(2\pi)^4}\frac{1}{(p^2+m_c^2)^n}
\ =\ \frac{1}{2^4\pi^2(n-1)(n-2)m_{c}^{2(n-2)}}\label{eq:mom}
\end{eqnarray}
and
\begin{displaymath}
\Tr\gamma_5(\sG)^3=2^5i\ \tr_c G\widetilde{G}G
=-2^3f_{abc}G^a\widetilde{G}^bG^c\ ,
\end{displaymath}
where $G\widetilde{G}G
=G_{\mu\nu}\widetilde{G}_{\nu\alpha}G_{\alpha\mu}$
and $f_{abc}$ is the structure constant of the color group.
As a result,
\begin{equation}
H_3(x)=-\frac{ig^3}{2^4\cdot 3\pi^2m_c^2}f_{abc}G^a\widetilde{G}^bG^c\ .
\label{eq:H3}
\end{equation}

We still have to calculate $H_2(x)$ 
which contains the gluon field strength $G_{\mu\nu}$ 
to the second power.
However, the calculation of $H_2(x)$ needs much more efforts 
owing to the noncommutativity of the operators, which is not negligible.
For the sake of the systematic momentum integration of $H_2(x)$,
one must transfer all operators containing $\cP_\mu$ 
to the left hand side in the trace.
Using the following expansion :
\begin{eqnarray}
\sG\frac{1}{\Klein}
&=&\frac{1}{\Klein}\sG+\frac{1}{\Klein}
[\cP^2,\sG]\frac{1}{\Klein}\nonumber\\
&=&\frac{1}{\Klein}\sG+\frac{1}{(\Klein)^2}[\cP^2,\sG]
+\frac{1}{(\Klein)^3}[\cP^2,[\cP^2,\sG]]\nonumber\\
&&+\frac{1}{(\Klein)^4}[\cP^2,[\cP^2,[\cP^2,\sG]]]+\cdots\ ,
\end{eqnarray}
we can rewrite $H_2(x)$ as
\begin{eqnarray}
H_2(x)
&=&
-\frac{ig^2m_c^2}{2}\bra{x}\Tr\gamma_5
\left\{\frac{1}{(\Klein)^3}(\sG)^2
+\frac{1}{(\Klein)^4}
\left(2[\cP^2,\sG]\sG+\sG[\cP^2,\sG]\right)\right.\nonumber\\
&&\left.+\frac{1}{(\Klein)^5}\left(3[\cP^2,[\cP^2,\sG]]\sG
+3[\cP^2,\sG]^2+\sG[\cP^2,[\cP^2,\sG]]\right)+\cdots\right\}\ket{x}
\nonumber\\
&\equiv&h_1(x)+h_2(x)+h_3(x)+\cdots
\ .\label{eq:12}
\end{eqnarray}
The calculation of the commutators is performed systematically.
For an arbitrary operator $\cQ ,$ 
which satisfies $[\cP_\mu, \cQ]=i\CovD_\mu\cQ$,
\begin{displaymath}
[\cP^2,\cQ]=\CovD^2\cQ+2i\cP_\mu \CovD_\mu\cQ\ .
\end{displaymath}
Repetitive use of this identity leads to
\begin{eqnarray}
&[\cP^2,\sG]&=\CovD^2\sG+2i\cP_\mu \CovD_\mu\sG\label{eq:com1}
\\
&[\cP^2,[\cP^2,\sG]]&=D^4\sG+2i\CovD_\nu G_{\nu\mu} 
\cdot \CovD_\mu \sG+2i \cP_\mu(\CovD_\mu \CovD^2+\CovD^2 \CovD_\mu)
\sG\label{eq:double}
\nonumber
\\
&&\;\;\;\;-4\cP_\nu G_{\nu\mu} \CovD_\mu \sG
-4 \cP_\nu \cP_\mu \CovD_\nu \CovD_\mu \sG\ .\label{eq:9}
\end{eqnarray}
In the following,
we assume the quasi-classical vacuum and 
the source $\rho$, which is defined by
$[\cP_\mu,G_{\mu\nu}]=i\CovD_\mu G_{\mu\nu}=\rho$,
is sufficiently small.
So, the second term in (\ref{eq:9}) may be negligible.
Since other higher commutators contribute in $H_2(x)$ as
at least third power of $G_{\mu\nu}$,
we can neglect the corresponding terms
which are denoted by dots in (\ref{eq:12}).

The momentum integration of each term can be performed 
easily since we need only both 
${\cal O}(G^2)$ and ${\cal O}(G^3)$ terms.
The calculation of the first term in $H_2(x)$, $h_1(x)$, is
somewhat technical.
By using the translational invariance of 
\begin{displaymath}
\bra{x}\Tr\gamma_5\frac{1}{\Klein}(\sG)^2\ket{x}
\end{displaymath}
with an arbitrary momentum $q_\mu$ 
and extracting all $q^2$-terms
as in Appendix,
$h_1(x)$ is given as 
\begin{equation}
h_1(x)=-\frac{ig^2m_c^2}{2}\bra{x}\Tr\gamma_5
\frac{1}{(\Klein)^3}(\sG)^2\ket{x}
=
\frac{ig^2}{2^4\pi^2}G^a\widetilde{G}^a+{\cal O}(G^4)\ .\label{eq:14}
\end{equation}
However, ${\cal O}(G^4)$ term
may be neglected for our purpose.
So we may regard the calculation of $h_1(x)$
as an integral over an ordinary momentum $p_\mu$,
as we have calculated in (\ref{eq:mom}).
The term $h_1(x)$ contributes to cancel the axial anomaly 
from the noninvariance of the fermionic measure.

The second term in $H_2(x)$, $h_2(x)$, is rewritten as
\begin{eqnarray}
h_2(x)&=&
-\frac{ig^2m_c^2}{2}\bra{x}\Tr\gamma_5
\frac{1}{(\Klein)^4}\left(2[\cP^2,\sG]\sG+\sG[\cP^2,\sG]\right)\ket{x}\\
\nonumber\\
&=&
-\frac{ig^2m_c^2}{2}\left\{\bra{x}\frac{1}{(\Klein)^4}\ket{x}
\Tr\gamma_5\sG\cdot \CovD^2\sG\right.\\
&&+6i
\left.\bra{x}\frac{1}{(\Klein)^4}\cP_\mu\ket{x}
\Tr\gamma_5\sG\cdot \CovD_\mu\sG
\right\}+({\rm Total\ derivative})\ .\nonumber
\end{eqnarray}
We can omit here also the terms 
which contain a single operator $\cP_\mu$.
The reason is that the matrix elements,
$\bra{x}(\Klein)^{-n}\cP_\mu\ket{x}$, must be denoted 
in terms of $G_{\mu\nu}$ and $\CovD_\mu$.
The first nonvanishing term 
which can give a contribution to this matrix element is $\CovD_\mu G^2$. 
It is clear that it leads to ${\cal O}(G^5)$, which we do not calculate here.

By using the Bianchi identity, it is easy to show that
\begin{equation}
\CovD^2G_{\mu\nu}=-ig\ [G_{\alpha\mu},G_{\alpha\nu}]
+\CovD_\mu \CovD_\alpha G_{\alpha\nu}-\CovD_\nu \CovD_\alpha G_{\alpha\mu}\ .
\end{equation}
where the second and third terms in the right-hand side may be neglected.
The reason is 
\begin{eqnarray*}
{\rm Tr}\gamma_5\sigma G \sigma_{\mu\nu}\CovD_\mu\CovD_\alpha G_{\alpha\nu}
&=&{\rm tr}_L \gamma_5\sigma_{\lambda\rho}\sigma_{\mu\nu}\cdot
   {\rm tr}_C G_{\lambda\rho}\CovD_\mu\CovD_\alpha G_{\alpha\nu}\\
&=&-4\varepsilon_{\lambda\rho\mu\nu}
   {\rm tr}_C G_{\lambda\rho}\CovD_\mu\CovD_\alpha G_{\alpha\nu}\\
&=&4\varepsilon_{\lambda\rho\mu\nu}
   {\rm tr}_C \CovD_\mu G_{\lambda\rho}\cdot\CovD_\alpha G_{\alpha\nu}
   + {\rm (Total\ derivative)}\\
&=&8{\rm tr}_C\CovD_\mu\widetilde{G}_{\mu\nu}\cdot\CovD_\alpha G_{\alpha\nu}
   + {\rm (Total\ derivative)}\\
&=&0+{\rm (Total\ derivative)},
\end{eqnarray*}
because of the Bianchi identity, $\CovD_\mu\widetilde{G}_{\mu\nu}=0$.
Then, $\CovD^2G_{\mu\nu}$ is the second power of $G$, and
\begin{equation}
\CovD^2 \sG=-2ig\sigma_{\mu\nu}G_{\alpha\mu}G_{\alpha\nu}\label{eq:D2sG}\ .
\end{equation}
The solution of $h_2(x)$ is obtained as
\begin{equation}
h_2(x)=\frac{ig^3}{2^4\cdot 3\pi^2m_c^2}f_{abc}G^a\widetilde{G}^bG^c
=-H_3(x)\ .
\end{equation}
Hence, $h_2(x)$  cancels with $H_3(x)$ in (\ref{eq:6}).
Thus, it will be only the term $h_3(x)$ in $H_2(x)$,
which contribute to the divergence of the axial current of charmed quarks.

Our remaining work is to calculate  only the third term in $H_2(x)$,
namely,
\begin{eqnarray}
h_3(x)
=-\frac{ig^2m_c^2}{2}\bra{x}\Tr\gamma_5\frac{1}{(\Klein)^5}
\left(3[\cP^2,[\cP^2,\sG]]\sG\right.\nonumber\\
\left.+3[\cP^2,\sG]^2+\sG[\cP^2,[\cP^2,\sG]]\right)&
\ket{x}\ .\label{eq:h3}
\end{eqnarray}
By using the expansion of the double commutator (\ref{eq:double}),
products of commutators and $\sG$
satisfy the relation,
\begin{eqnarray}
[\cP^2,[\cP^2,\sG]]\sG&=&-[\cP^2,\sG]^2
=\sG[\cP^2,[\cP^2,\sG]]\nonumber\\
&=&-4\cP_\nu\cP_\mu\sG \CovD_\nu \CovD_\mu\sG
\end{eqnarray}
except both higher order terms and nonessential total derivatives 
which are not needed in our purpose.
So the first term in (\ref{eq:h3}) cancels with the second term.
Hence, $h_3(x)$ is reduced to
\begin{equation}
h_3(x)=2ig^2m_c^2\Tr\bra{x}\frac{1}{(\Klein)^5}\cP_\nu\cP_\mu\ket{x}
\gamma_5\sG \CovD_\nu \CovD_\mu\sG\ .\label{eq:h3b}
\end{equation}
Here, owing to the algebraic relation (\ref{eq:1}), 
an extra $G_{\mu\nu}$ could appear as
$\cP_\nu\cP_\mu=
\frac{1}{2}\{\cP_\nu,\cP_\mu\}+igG_{\nu\mu}$.
However, 
since there is the ${\cal O}(G^3)$ content in the trace,
we can neglect the noncommutativity
in the matrix element of (\ref{eq:h3b}).
Then the operator $\cP_\mu$ can be replaced by $p_\mu$
as we performed in the calculation of $H_3(x)$ before :
\begin{eqnarray}
\bra{x}\left.\frac{1}{(\Klein)^n}\cP_\nu\cP_\mu\right|_{A=0}\ket{x}
&=&\int\frac{d^4p}{(2\pi)^4}\frac{p_\nu p_\mu}{(p^2+m_c^2)^n}\nonumber\\
&=&\frac{\delta_{\mu\nu}}{2^5\pi^2(n-1)(n-2)(n-3)m_c^{2(n-3)}}\;\ ,
\label{eq:PP}
\end{eqnarray}
Substituting Eqs.(\ref{eq:D2sG}) and (\ref{eq:PP}) for (\ref{eq:h3b}), 
we obtain 
\begin{equation}
h_3(x)=-\frac{ig^3}{2^5 \cdot 3\pi^2 m_c^2}f_{abc}G^a\widetilde{G}^bG^c\ .
\end{equation}
Hence,
\begin{eqnarray}
H_2(x)&=&h_1(x)+h_2(x)+h_3(x)\nonumber\\
&=&\frac{ig^2}{2^4\pi^2}G^a\widetilde{G}^a
+\frac{ig^3}{2^5\cdot 3\pi^2 m_c^2}f_{abc}G^a\widetilde{G}^bG^c\ .
\end{eqnarray}
Finally, since $h_2(x)$ cancels $H_3(x)$, we get
\begin{eqnarray}
H(x)&=&h_1(x)+h_3(x)\nonumber\\
&=&\frac{ig^2}{2^4\pi^2}G^a\widetilde{G}^a
-\frac{ig^3}{2^5\cdot 3\pi^2 m_c^2}f_{abc}G^a\widetilde{G}^bG^c\ .
\end{eqnarray}
As  expected, the first term in $H(x)$ cancels with the first term in 
 \re{divc}, which is the contribution from noninvariance of the measure
and the rest part leads to the divergence of the $c$-quark 
axial current in the form
\be
\langle\partial_{\mu}c^{\dagger}(x)\gamma_{\mu}\gamma_{5}c(x)\rangle 
\,=\,\,-\, 
\frac{i g^{3}}{2^{5}3\pi^{2}m_{c}^{2}}f_{abc}G^{a}\tilde G^{b} G^{c} .
\lab{divc-1}
\ee
We would like to stress  that our answer for 
$\langle\partial_{\mu}c^{\dagger}(x)\gamma_{\mu}\gamma_{5}c(x)\rangle$
is $6$ times less
than that of Halperin and Zhitnitsky \ci{HZ}.

\section{Matrix elements of gluonic operators in Effective Action approach}

We face now the problem of the calculations of  various matrix elements
of the gluonic operators like those in \re{div} and \re{divc-1}.  

First we rederive the Diakonov-Petrov (DP) Effective Action \ci{DP} as it was
suggested in \ci{SM}.
It is natural to choose the singular gauge for the instantons in describing
many instanton effects in the propagation of the quarks. In the case of
a small packing parameter  
it is possible to do the following sum ansatz for the 
background instanton field:
\be
A_{ \mu}(x)= \sum_{+}^{N_+} A_{+ , \mu}(x; \xi_{+}) + 
\sum_{-}^{N_-} A_{-, \mu}(x; \xi_{-})\;\;, \;\; 
(\xi_{\pm}=(z_{\pm}, U_{\pm},\rho_{\pm})),
\ee
where $z_i$, $U_i$ and $\rho_i$ are the position, orientation and size of the
$i$-th instanton.
The canonical partition function of the
 $N_+$instantons and $N_-$ anti-instantons can be schematically
written as
\be
Z_{N_+ , N_-} = \int {\rm det}_N \exp( - V_g )\prod_{i}^{N_+ , N_-} 
d^4 z_i dU_i dn(\rho_i),
\ee                                                      
where $V_g$ is the instanton-(anti)instanton interaction potential generated
by the gluon field action and ${\rm det}_N$ is a quark determinant in the instanton 
field. The main assumption of the instanton model is  that
$V_g$ is repulsive  at small distances between instanton and anti-instanton.
This should 
provide the stabilization of the instanton sizes and of the inter-instanton
distances. We mainly deal with ${\rm det}_N ,$ 
which describes the influence of light quarks.

Lee\&Bardeen \ci{LB79} (LB) calculated the quark propagator in a more 
sophisticated approximation than DP, finding
\be
{\rm det}_N=\det B, \,\,
B_{ij}= im\delta_{ij} + a_{ji}, 
\lab{B}
\ee
where $a_{ij}$ is the overlapping matrix element of the quark zero-modes 
$\Phi_{\pm , 0} $ generated by instantons\footnote{$\Phi_{\pm , 0} $
is the solution of the Dirac equation $\Fs{\cP}\Phi_{\pm , 0}\,=\,0$
in the instanton(anti-instanton) field 
 $A_{\pm , \mu}(x; \xi_{\pm})$.}.
This matrix element is nonzero only between instantons and anti-instantons
(and vice versa) due to the chiral factor in $\Phi_{\pm , 0} $ , i. e.,
\be
a_{-+}=-\bra{\Phi_{- , 0} } i\hat\partial \ket{\Phi_{+ , 0} } .
\lab{a}
\ee
The overlap of the quark zero-modes causes quarks 
jumping from one instanton to another  during their propagation.

Eq. \re{B} implies that 
for $N_+\not=N_-$ ${\rm det}_N\sim m^{|N_+ - N_-|}$, so the fluctuations of  
$|N_{+}-N_{-}|$  are strongly suppressed due to the presence of light quarks. 
Therefore we assume $N_{+}=N_{-}=N/2 .$

Let us rewrite the ${\rm det}_N$ following the idea  suggested
in \ci{Tokarev}.
First, by introducing  the Grassmanian $(N_{+},N_{-})$ vector 
$$\Omega=(u_{1}...u_{N_{+}}, v_{1}...v_{N_{-}})$$
and
$$\bar\Omega=(\bar u_{1}...\bar u_{N_{+}}, \bar v_{1}...\bar v_{N_{-}})$$
we can rewrite 
\be
{\rm det}_N = \int d\Omega d\bar\Omega \exp (\bar\Omega B \Omega ) ,
\ee
where
\be
 \bar\Omega B \Omega  = \bar\Omega (im + a^{T}) \Omega =
i\sum_{+}m\bar u_{+}u_{+} + i \sum_{-}m\bar v_{-}v_{-} 
+ \sum_{+-} (\bar u_{+}v_{-}a_{-+} + \bar v_{-}u_{+}a_{+-}) .
\ee
The product  $\bar u_{+}v_{-}a_{-+}$ can be rewritten in the form
\be
\bar u_{+}v_{-}a_{-+}=  
- \bra{i\hat\partial \Phi_{- , 0} v_{-}} 
(i\hat\partial)^{-1} \ket{i\hat\partial \Phi_{+ , 0} \bar u_{+}} .
\ee
The next step is to introduce   $N_{+},N_{-}$ sources 
$\eta=(\eta_{+}, \eta_{-})$
and $N_{-},N_{+}$ sources $\bar\eta=(\bar\eta_{-}, \bar\eta_{+})$
defined as:
\bea
\bar\eta_{-}=\bra{i\hat\partial \Phi_{- , 0} v_{-}} &,&
\bar\eta_{+}=\bra{i\hat\partial \Phi_{+ , 0} u_{+}}  \nonumber\\ 
\eta_{+}=\ket{i\hat\partial \Phi_{+ , 0} \bar u_{+}} &,&
\eta_{-}=\ket{i\hat\partial \Phi_{- , 0} \bar v_{-}}\;\;.\nonumber
\eea
Then $\exp(\bar\Omega a^{T} \Omega)$ can be rewritten as
\bea
\lefteqn{\exp(\bar\Omega a^{T} \Omega )}\nonumber\\
&=&\exp\int \left(-\bar\eta (i\hat\partial)^{-1} \eta \right)\nonumber\\
&=& 
\left(\det(i\hat\partial)\right)^{-1} \int D\psi D\psi^{\dagger}
\exp\int dx\left(\psi^{\dagger} (x)
i\hat\partial\psi (x)-\bar\eta (x)\psi (x)+\psi^{\dagger} (x)\eta (x)\right)
\eea
Integrating  over Grassmanian variables $\Omega$ and $\bar\Omega$
and taking into account  the $N_{f}$ flavors 
${\rm det}_N = \prod_{f}\det B_{f}$
this  provides the fermionized representation of the Lee\& Bardeen's result
for ${\rm det}_N$ in the form:
\be
\ba  \Ds
{\rm det}_N = \int D\psi D\psi^{\dagger} \exp\left(\int d^4 x
\sum_{f}\psi_{f}^{\dagger}i\hat\partial \psi_{f}\right)     \\   \Ds
\times \prod_{f}\left\{\prod_{+}^{N_{+}}
\left(im_{f} + V_{+}[\psi_{f}^{\dagger} ,\psi_{f}]\right)
\prod_{-}^{N_{-}}
\left(im_{f}+V_{-}[\psi_{f}^{\dagger},\psi_{f}]\right)\right\}\; ,
\label{part-func}
\ea
\ee
where 
\be
V_{\pm}[\psi_{f}^{\dagger} ,\psi_{f}]= 
\int d^4 x \left(\psi_{f}^{\dagger} (x) i\hat\partial
\Phi_{\pm , 0} (x; \xi_{\pm})\right)
\int d^4 y 
\left(\Phi_{\pm , 0} ^\dagger (y; \xi_{\pm} )  
i\hat\partial \psi_{f} (y)\right).
\ee
Eq. \re{part-func} coincides with the ansatz 
for the fixed $N$ partition function
postulated by DP, except for the sign in front of 
$V_{\pm}$.  Keeping in mind the low density of the instanton media, which 
allows independent averaging over positions and orientations 
of the instantons, Eq. \re{part-func} leads to the partition function

\be
 Z_N = \int D\psi D\psi^\dagger \exp  \left(\int d^4 x \, \psi^\dagger 
i \hat\partial  \psi \right)  \,  W_{+}^{N_+}  \, W_{-}^{N_-}, 
\lab{Z_NW}
\ee
where
\be\ba\Ds
W_\pm =\int d^4 \xi_{\pm}\prod_{f}\left(V_{\pm}[\psi_{f}^{\dagger} \psi_{f}]+
\ i m_{f}\right)
\\ \Ds
=(-i)^{N_{f}}\left(  \frac{4\pi^2 \rho^2}{N_c} \right)^{N_f}
\int \frac{d^4 z}{V} 
{\rm det}_{f}\left(i J_\pm (z) - \frac{m ~ N_c}{4\pi^2 \rho^2}\right)
\ea\ee

and

\be
J_\pm (z)_{fg} = \int \frac {d^4 kd^4 l}{(2\pi )^8 } 
\ e^{-i(k - l)z}
\, F(k^2) F(l^2) \, \psi^\dagger_f (k) \half (1 \pm \gamma_5 ) \psi_g (l) .
\lab{J_pm}
\ee
The form factor $F$ is related  to the zero--mode wave function 
 in momentum space $\Phi_\pm (k; \xi_{\pm}) $ and is equal to
\be
F(k^2) = - t \frac{d}{dt} \left[ I_0 (t) K_0 (t) - I_1 (t) K_1 (t)
\right]\;\;, \;\; t =\frac{1}{2} \sqrt{k^2} \rho.
\ee
\\ \\
\underline{ Correlators in the $DP$ effective action}

In quasiclassical (saddle point) approximation any gluon operator receives its 
main contribution from instanton background. 
In the following   the operator $g^2 G\tilde G(x)$  will be considered for the
illustration of the method.
Owing to the low density of the instanton medium, it is possible to neglect  
the overlap of the fields of different instantons. In that case,  
the matrix element of $g^2G\tilde G(x)$ with any other 
quark operator $Q$ is
\be \ba\Ds
\langle {g^2}G\tilde G(x) Q \rangle_N = 
Z_{N}^{-1} \int D\psi D\psi^\dagger \exp\left(\int d^4 x \psi^\dagger i 
\hat\partial \psi\right)   
\\ \\
 \times
\left\{ N_{+} \left( W_{G\tilde G +} (x) Q \right) W_{+}^{N_+ - 1}   
W_{-}^{N_-}  +  N_{-} \left( W_{G\tilde G -} (x) Q \right) W_{+}^{N_+ }   
W_{-}^{N_- - 1} \right\} ,
\lab{GtildeGQ1}     
\ea \ee
where
\be
W_{G\tilde G \pm} = \pm\left(  \frac{4\pi^2 \rho^2}{N_c} \right)^{N_f}
\int \frac{d^4 z}{V}\, f_{2}(x-z) \, 
{\rm det}_{f}\left( J_\pm (z) + i\frac{m ~ N_c}{4\pi^2 \rho^2}\right) .
\lab{W_GtildeGQ}
\ee 
and
$f_{2}(x-z)$ is defined as 
\be
(g^{2}G\tilde G(x))_{\pm} = \pm f_{2}(x-z)=
\pm\frac{192 \rho^4}{\left[ \rho^2 + (x - z)^2 \right]^4}
\lab{GtildeG}   
\ee
It is useful to introduce the external field $\kappa_{2(3)} (x)$, coupled 
respectively to $g^2 G\tilde G$ ($g^3 f_{abc}G^{a}\tilde G^{b} G^{c}$).
Starting from \re{GtildeGQ1} and \re{W_GtildeGQ}, we find the partition function 
$\hat Z[\kappa_{2(3)}]$  
describing mesons\ci{MK} in presence of such external field:
\be
\hat Z[\kappa_{2(3)}]
=\int D\Phi_{+}D\Phi_{-}\exp\left(-W[\Phi_+,\Phi_-]\right),  
\label{intz}
\ee
where
\be\ba\Ds
W[\Phi_+,\Phi_-] = \int d^4 x (w_a + w_b -w_c),
\\ \\ \Ds
w_a = (N_{f}-1)\frac{N}{2V}\left(\prod_{f}M_{f}^{-1}\det \Phi_{+}\right)
^{(N_{f}-1)^{-1}}
\ + \;\left(\Phi_{+} \rightarrow \Phi_{-}\right)\;, 
\\ \\ \Ds
\ w_b = \frac{N_c}{4\pi^{2}\rho^{2}}\Tr\{m (\Phi_{+}+\Phi_{-})\},
\\ \\ \Ds
w_c = \sum_{f} \Tr \ln \frac{i \hat \partial  +i F^2 ~ (\Phi_+ \beta_+ + 
\Phi_- \beta_-)} {i \hat \partial + i m_f},
\\ \\ \Ds
\beta_{\pm} = 
\left[ \left( 1 \pm ( \kappa_{2(3)}  f_{2(3)}) \right)^{N_{f}^{-1}} \right]
\frac {1}{2}(1 \pm \gamma_5).
\\
\lab{Zkappa2}    
\ea
\ee
The two remarkable formulas
\be
 (ab)^N = \int d\lambda \exp (N ln \frac {aN} {\lambda} - N +\lambda b )
\,\,(N >>1).
\lab{ab^N}
\ee
and
\be
\exp (\lambda \det [i A] ) =
\int d\Phi \exp\left[ - (N_f - 1) \lambda^{-\frac{1}{N_f - 1}}
(\det\Phi )^{\frac{1}{N_f - 1}} + i tr (\Phi A) \right] 
\lab{expA} 
\ee
have been used here.
It is possible to check these formulas by the saddle point approximation  
of the integrals.  They were proposed in \ci{DPW}
and we followed this approach.

The saddle point  of the integral \re{intz} is located at 
$(\Phi_{\pm})_{fg} = M_{f}\delta_{fg}$, a self-consistency condition for the 
effective quark mass, i. e.,
\be
4 N_c V \int \frac{d^4 k}{(2\pi )^4} 
\frac{M_{f}^{2} F^4 (k^2)}{M_{f}^{2} F^4 (k^2) + k^2}
=  N  + \frac{m_{f}M_{f}VN_{c}}{2\pi^{2}\rho^{2}},
\lab{selfconsist}  
\ee
being imposed, which describes also the shift of the effective
mass of the quark $M_f$ due to current mass $m_f$. In the following we will
neglect  $m_f$, since this model fails to reproduce properly the low-energy
theorems beyond chiral limit \ci{SM}.

The solution of a self-consistency  equation \re{selfconsist} in chiral limit 
correspond to $M_{0} \,=\, 340\, {\rm MeV}$ 
assuming  the parameters $\rho$ and $R$, which are the values given in
\re{rho,R}.

The action $W[\Phi_+,\Phi_-]$ in \re{Zkappa2} has imaginary part, in general,
 which is reduced to Wess-Zumino term in long-wave limit 
($k\,\ll\, M$, where $k$ is a mesons momentum).

\section{Virtual charm in $ B \rightarrow K \eta' $ decay}

We apply the Effective Action \re{Zkappa2} and  the formula for the divergence 
of the c-quark axial current\re{divc-1} to the calculation of  
$f_{\eta '}^{(c)}$ and compare with  the analogous
quantity $f_{\eta '}^{(u)}$, which is defined in the similar way 
as $f_{\eta '}^{(c)}$ in\re{fc}. These quantities certainly are defined  in
Minkowski space.  The eqs.  \re{div} and \re{divc-1} may be easily 
translated from Euclidean to Minkowski space accordingly 
the above-given prescription. For instance
 $( (G^{a}\tilde G^{a})_E \rightarrow (G^{a}\tilde G^{a})_M)$  and
 $( (f_{abc}G^{a}\tilde G^{b} G^{c})_E 
 \rightarrow\,-\, (f_{abc}G^{a}\tilde G^{b} G^{c})_M),$ which lead to   
\be
m_{\eta '}^{2}f_{\eta '}^{(u)}\,=\,
<0|\frac{g^2}{16\pi^2} (G^{a}\tilde G^{a})_M |\eta ' > .
\lab{fu}
\ee
and
\be
m_{\eta '}^{2}f_{\eta '}^{(c)}\,=\,-\,<0| \frac{g^3}{2^{5}3\pi^{2}m_{c}^{2}}
(f_{abc}G^{a}\tilde G^{b} G^{c})_M |\eta ' >
\lab{fc-1}
\ee
The phenomenological way of the estimation of the $f_{\eta '}^{(u)}$ 
is the application of  the 
QCD+QED axial anomaly equation together with data on 
$\eta ' \rightarrow 2 \gamma$
decay leads to 
\be
f_{\eta '}^{(u)} \,=\, 63.6\, {\rm MeV},
\lab{fu-1}
\ee
which was used in \ci{ACGK97}.
On the other hand the calculation of the matrix elements \re{fu}, \re{fc-1}
may be reduced to the calculation of the correlators 
$$<G^{a}\tilde G^{a}(x)\,G^{a}\tilde G^{a}(y)>,\,\,\, 
<f_{abc}G^{a}\tilde G^{b} G^{c}(x)\,G^{a}\tilde G^{a}(y)>$$
respectively. The calculation of the correlators are naturally performed 
in Euclidean space.
These correlators  have almost the same dependence on the 
large relative distances
$|x-y|$ and the ratio of these correlators become almost constant, 
at least in  DP chiral quark model \ci{DP}. 

In the Effective Action approach, the abovementioned gluonic operators 
vertices in 
the correlators are generated by differentiation of the Effective Action 
\re{Zkappa2} over $\kappa_{2(3)}$, which leads to the vertices
\be
i\,f_{2(3)}\,F^{2}
\,N_{f}^{-1}\,\gamma_5 \frac{1}{2}
((\Phi_{+}+\Phi_{-})\,+\,(\Phi_{+}-\Phi_{-})\gamma_{5})
\lab{vertices1}
\ee
So,
 the calculations of $f_{\eta '}^{(c)}$
and also $f_{\eta '}^{(u)}$ may be reduced to the calculations of the similar 
two-point
$\gamma_5$-singlet correlators with additional form-factor 
$f_3 (q)$ or $f_2 (q)$ \re{GtildeG}, which are simply 
the momentum representation of the instanton contribution to the 
operators $g^{3}f_{abc}G^{a}\tilde G^{b} G^{c}$ and $g^{2} G^{a}\tilde G^{a}$ 
(see Eq. \re{GtildeG}), respectively. They are defined as
\bea
f_3 (q)\,=
\, -1536\, \rho^{6}\, \int d^{4}x \,\exp(iqx)\, (\rho^{2} + x^{2})^{-6}, 
\nonumber\\
f_2 (q)\,=
\, 192\, \rho^{4}\, \int d^{4}x \,\exp (iqx)\, (\rho^{2} + x^{2})^{-4}, 
\lab{f2f3}
\eea
and
\be
f_2 (0)\,=\, 32\pi^{2}, \,\, f_3 (0)\,=\, - \frac{12}{5\rho^{2}}f_2 (0)
\lab{f2f3(0)}
\ee
It is easy to reduce the integrals in \re{f2f3} to the Bessel functions as
\be
 \int d^{4}x \,\exp(iqx)\, (\rho^{2} + x^{2})^{-n} = \frac{\pi^{2}}{(n-1)!}
 \left(\frac{q^{2}}{\rho^{2}}\right)^{(n-3)/2}
 K_{n-3}\left((q^{2}\rho^{2})^{1/2}\right)\;.
\lab{bessel}
\ee
With
 the Effective Action  \re{Zkappa2} we  calculate everything  in Euclidean
space with  further  analytical continuation to the Minkowski region. 

On the other hand the ratio of abovementioned correlators 
in the Effective Action 
approach  equal the ratio of form-factors 
$f_3 (q)$ and $f_2 (q)$. This ratio has a weak dependence on the
argument (like 10\% on the scale $q \sim \rho^{-1}$). 
As a result, this  provides the possibility
to calculate the ratio $f_{\eta '}^{(c)}/f_{\eta '}^{(u)}$
 at small Euclidean $q^2 $ within this accuracy.  

Now it is clear that with the Effective Action \re{Zkappa2} 
 the 
matrix element in \re{fc-1}  may be reduced to 
the calculation of the matrix element in  \re{fu} with an additional
factor $-\frac{12}{5\rho^{2}}$. 
So, the ratio of Eqs. \re{fc-1} and \re{fu}
is equal in this model to:
\be
\frac{f_{\eta '}^{(c)}}{f_{\eta '}^{(u)}} \,=\,-\frac{12}{5\rho^{2}}
\frac{1}{6m_{c}^{2}} \,\sim\, -0.1,
\lab{ratio}
\ee
where  the value for $\rho$ is given in \re{rho,R}. 
By taking into account the estimation \re{fu-1}
(we use $m_c(\mu\simeq m_c)\simeq 1.25\, {\rm GeV}$  on the scale 
$\mu\simeq m_c$ for the numerical estimates), 
we find
\be
f_{\eta '}^{(c)} \,=\,-\, 6\, {\rm MeV}.
\lab{fc-2}
\ee
This number is close to the one of \ci{AG97},
$|f_{\eta '}^{(c)}| \,=\, 5.8\, {\rm MeV}$
and   the sign  and the order of the value coincide with the estimations of 
\ci{ACGK97,FK98,P}.

Recently, Shuryak and Zhitnitsky \ci{SZ} performed direct numerical
evaluations of the various
correlators of the operators $g^{2} G^{a}\tilde G^{a}$, 
$g^{3}f_{abc}G^{a}\tilde G^{b} G^{c}$ in the Interacting Instanton Liquid
Model(IILM).
Their calculations lead to:
\be
\bra{0}g^{2} G^{a}\tilde G^{a}\ket{\eta '} \,=\,7\, {\rm GeV}^3 ,
\lab{SZGtildeG}
\ee
(which leads to $f_{\eta '}^{(u)} \,=\, 48.3\, {\rm MeV}$)
and 
\be
\frac{|\bra{0}g^{3}f_{abc}G^{a}\tilde G^{b} G^{c}\ket{\eta ' }|}
{|\bra{0}g^{2} G^{a}\tilde G^{a}\ket{\eta '}|}
\approx  (1.5\sim 2.2)\,{\rm GeV}^2 .
\lab{sz}
\ee
 The later is somewhat large than their simple estimate for this ratio
of matrix elements  
\be 
{12\over 5} \left\langle {1\over \rho^2} \right\rangle 
\approx  (1\sim 1.5)\, {\rm GeV}^2
\lab{est-sz}
\ee
With the use of \re{sz} and  \re{divc-1} we arrive at
\be
\frac{f_{\eta '}^{(c)}}{f_{\eta '}^{(u)}} \,=\,- 0.17\sim - 0.25 .
\lab{ratio1}
\ee
This ratio  gives  $f_{\eta '}^{(c)}\,=\,- 8.2\sim -12.3\, {\rm MeV}$ 
at the scale of the size of the instanton $\mu\approx \rho^{-1} .$
The abovementioned experimental numbers \re{etapkpm} are given at
 the scale $\mu\approx m_c$, which is
 different from the scale of  this instanton calculation.
 The account of the   anomalous dimension of the 
$g^3G\tilde{G}G$ operator \ci{M84}  leads to correction  \ci{SZ}  
\be
\label{correction}
f_{\eta '}^{(c)} (\mu \simeq m_c)
\simeq 1.5 f_{\eta '}^{(c)} (\mu \simeq \rho^{-1}) .
\ee
 The account of this scale  factor leads to
\be
\label{fc-4}
f_{\eta '}^{(c)} (\mu \simeq m_c)\,=\,- 12.3\sim -18.4\, {\rm MeV} .
\ee
Hence, using \re{sz}, the result of more sophisticated calculations 
of Shuryak and Zhitnitsky, we get
 the number \re{fc-4} which  is 2-3 times 
larger than our simple estimation \re{fc-2}.

These numbers \re{fc-2}, \re{fc-4} are in  agreement with the
phenomenological bounds
\cite{FK97,P} and almost in agreement in the sign and the 
value with \ci{ACGK97,FK98} but 
 six-ten times less than the estimations given by  
 \ci{HZ}(see \re{HZestimation}).

By using  the numerical analysis of the branching ratio for 
$B^\pm \rightarrow \eta^{\prime}  K^\pm$
 given at
 \ci{AG97} (Fig.17 of \ci{AG97} ) we expect that the value of 
 $f_{\eta '}^{(c)}$ given in  \re{fc-4} may provide a more satisfactory
 explanation of the experimental data.

\section{Virtual charm in polarized DIS}

From the definition Eq. \re{def1} and from Eqs. \re{div}, \re{divc-1} 
it is easy to find 
\bea
\Delta \Sigma \,2m_{N}\, \bar{p} i\gamma_5   p\, = \,
\langle p,s \vert  \frac{N_{f}g^2}{16\pi^2} G\tilde G
 \vert p,s\rangle ,
 \\
\Delta c \,2m_{N}\, \bar{p} i\gamma_5   p\, = \,-\,
\langle p,s \vert 
 \frac{i g^{3}}{2^{5}3\pi^{2}m_{c}^{2}}f_{abc}G^{a}\tilde G^{b} G^{c}
 \vert p,s\rangle 
\lab{deltas,deltac}
\eea
(These formulas are defined in Minkowski space).
As it was mentioned before $\Delta \Sigma\,=\, 1$ 
in the constituent quark model.
On the other hand it was shown that  Skyrme soliton model gives zero for
 this quantity.
Chiral quark-soliton model (see recent review of this model \ci{D98})
interpolate between these  constituent quark
and Skyrme models and give  the value of 
$\Delta \Sigma$  in the range \ci{WY90,BPG93}
\be
\Delta \Sigma \,=\,0.3 \,\sim \, 0.5.
\lab{DS}
\ee 
The calculation of the same quantity by QCD sum rules approach give
the same answer (see the review \ci{Ioffe98}). 

Chiral quark-soliton model is essentially based on the Effective Action like 
 \re{Zkappa2}, where all the degrees of freedom are frozen except constituent
quarks and pions. In that case, by restoring of the quark degrees of freedom 
in \re{Zkappa2}, we find 
\be
\hat Z_{N}[\kappa_{2(3)}]=  
\int D\psi^{\dagger}D\psi DU \exp \int \psi^{\dagger}
\left\{i \hat \partial  +i F^2 M U^{\gamma_5}
\left( 1 + \gamma_{5}( \kappa_{2(3)}  f)\right)^{N_{f}^{-1}}\right\}\psi ,
\label{Z_N}
\ee
where
$U^{\gamma_5} \,=\, \frac{1}{2M}
((\Phi_{+}+\Phi_{-})\,+\,(\Phi_{+}-\Phi_{-})\gamma_{5})$,
$\Phi_{\pm} = M \exp(\pm i\phi )$ and
the usual decomposition for the  pions
$\phi = \sum_{1}^{3}\tau_{i}\phi_{i}$   
may be used. 

The mass of nucleon is calculated from Euclidean 
large-distance asymptotes of 
two-point correlator of the composite quark operators $\Gamma_{N}(x)$
with nucleon quantum numbers and certainly by using $\hat Z_{N}[0]$.
 
The nucleon mass  receive the contributions  
from the $N_c$ products of the constituent quark propagators in the
external pion field and the polarization of the constituent quarks vacuum by
this field(effective action for the pions) 
integrated over this field. The saddle-point condition in this path
integral means topologically nontrivial pion field $\phi_{s}$
like famous skyrmion (see for example \ci{D98}).

The calculation of the nucleon matrix elements of the any combination of the
gluonic operators  $g^{2}G^{a}\tilde G^{a}$ and 
$g^{3}f_{abc}G^{a}\tilde G^{b} G^{c}$ 
 may be reduced to the differentiation over $\kappa_{2(3)}$ of 
the two-point correlator of the composite quark operators $\Gamma_{N}(x)$
calculated now by using  $\hat Z_{N}[\kappa_{2(3)}]$.

This is the way to calculate  $\Delta \Sigma $ and $\Delta c ,$ 
which are essentially reduced  to the nucleon matrix elements of
the gluonic operators 
$g^{2}G^{a}\tilde G^{a}$
and $g^{3}f_{abc}G^{a}\tilde G^{b} G^{c}$ respectively.
It is clear, that these operators lead to the analogous \re{vertices1}
vertices
\be
i\,f_{2(3)}\,M\,\,F^{2}\,U^{\gamma_5}
\,N_{f}^{-1}\,\gamma_5 .
\lab {vertices2}
\ee
The  form-factors 
$f_2 $ and $f_3$ are defined in \re{f2f3}. 
In the present case we need these form-factors at $q^{2}=0$.

On the very general ground, it is possible to prove that the calculations
of $  \Delta \Sigma  $ with using 
$\hat Z_{N}[\kappa_{2}]$ leads exactly  the result of 
Wakamatsu and Yoshiki \ci{WY90}. In that paper authors calculated 
the nucleon matrix element of the operator 
$ u^{\dagger}\gamma_{\mu}\gamma_{5}u \,+\, d^{\dagger}\gamma_{\mu}\gamma_{5}d$
by using the Effective Action like $\hat Z_{N}[0]$.

Let us  change variables in $\hat Z_{N}[\kappa_{2}]$, Eq. \re{Z_N},
accordingly:
\be 
\psi^{'}\,=\,\exp\left(\alpha (z)\gamma_{5}\right)\psi ,\,\,\,\,
\psi^{\dagger '}\,=\,\psi^{\dagger}\exp\left(\alpha (z)\gamma_{5}\right) .
\lab{transf}
\ee
 If we choose 
 $$\alpha \,=\,1/2( \kappa_{2}  f_{2})^{N_{f}^{-1}}$$
 then most important $O(\kappa_{2} )$ term 
 in $\hat Z_{N}[\kappa_{2}]$will appear now in an other form as:
\begin{eqnarray}
\hat Z_{N}[\kappa_{2}]&=&\int\,D\psi^{\dagger '}\,D\psi ^{'}\,J\,DU
\nonumber\\
&&\times \exp \int\,dz\,\psi^{\dagger '}(z)
\left\{i \hat \partial \,
+\frac{1}{2}\,\gamma_{\mu}\gamma_{5}\,\left(\int\,dx\,\kappa_{2}(x)
\partial_{z,\mu} f_{2}(x-z)\right)^{N_{f}^{-1}}
+i F^2 M U^{\gamma_5}\right\}\psi^{'}(z)
\label{Z_N-transf}
\end{eqnarray}
where  $J$ is a Jacobian of the transformations of the measure. 
Such type  of Jacobians
is responsible for the chiral anomaly, in general.
In the present case we are dealing with  
the action in \re{Zkappa2}, \re{Z_N}, which  has a imaginary part, since
Dirac operator 
$$D\,=\,i \hat \partial \,+\,i F^2 M U^{\gamma_5}$$
is not hermitian.
To calculate the Jacobian $J$ we must properly define the measure.

First of all we must choose the total systems of the eigenfunctions of some
hermitian operators. 
It is natural to take the operators:
\be
D^{+}D\,=\,-\,\partial^{2} \,+\, F^{4} M^{2}\, - 
\, F^{2} M(\hat \partial U^{\gamma_5}),\,\,
 DD^{+}\,=\,-\,\partial^{2} \,+\, F^{4} M^{2}\, + 
\, F^{2} M(\hat \partial U^{\gamma_{5}+}),
\lab{DD^+}
\ee
and define a set of the eigenfunctions:
\be
D^{+}D\,\phi_{n}\,=\,\lambda_{n}^{2}\,\phi_{n}, \,\,\,
 DD^{+}\,\Phi_{n}\,=\lambda_{n}^{2}\,\Phi_{n}.
\lab{phi_n}
\ee
It is easy to show that:
$$D\,\phi_{n}\,=\,\eta_{n}\,\Phi_{n},$$
where $|\eta_{n}|\,=\,|\lambda_{n}|.$

With this basis we may  expand:
\be
\psi (x)\,=\,\sum_{n}\,a_{n}\,\phi_{n},\,\,
\psi^{\dagger} (x)\,=\,\sum_{n}\,b^{\dagger}_{n}\,\Phi^{+}_{n}, 
\lab{expansionpsi}
\ee
where $a_{n}$ and $b^{\dagger}_{n}$ are Grassmanian numbers.
So, the measure may be defined as 
$$ D\psi \,=\,\prod_{n}\,da_{n},\,\,
D\psi^{\dagger}\,=\,\prod_{n}\,db^{\dagger}_{n}.$$
The Jacobian $J$ then is given by
\be
J \,=\, \exp (\int\,dx\,\alpha (x)\,A(x)).
\lab{J}
\ee
According to Fujikawa \ci{Fujikawa}, the regularized expression for the 
anomaly $A(x)$ is given by 
\be\ba\Ds
A(x)\,=\,\sum_{n}\phi^{+}_{n}(x)\gamma_{5}
\exp(-\frac{D^{+}D}{\mu^2})\phi_{n}(x) \,+\,\sum_{n}\Phi^{+}_{n}(x)\gamma_{5}
\exp(-\frac{DD^{+}}{\mu^2})\Phi_{n}(x)
\\ \Ds
=\,\lim_{\mu \to \infty} tr\int\,\frac{d^{4}k}{(2\pi )^{4}}
\exp(-ikx)\gamma_{5}\left\{\exp(-\frac{D^{+}D}{\mu^2})\,+\,
\exp(-\frac{DD^{+}}{\mu^2})\right\}\exp(ikx) .
\lab{anomaly}
\ea\ee
Redefining the variable of the integration in \re{anomaly} $k \, \to \, k/\mu$ 
we arrive at:
\be
A(x)\,=\,tr\int\,\frac{d^{4}k}{(2\pi )^{4}}\exp(-k^{2})
\gamma_{5}[(F^{2} M(\hat \partial U^{\gamma_5}))^{2}
\,+\,(F^{2} M(\hat \partial U^{\gamma_{5} +}))^{2}]\,=\, 0,
\lab{A} 
\ee
due to the trace over Dirac matrices.
So, Jacobian $J$ is equal to one 
and now it is absolutely clear that the calculations with Eqs. 
\re{Z_N-transf}, \re{Z_N} leads to the same result 
in \ci{WY90,BPG93}. 

The $SU(3)$ extensions \ci{BPG93} of the model \re{Z_N} will leads to the same
result for $\Delta \Sigma ,$ since the valence $u,\, d$ quarks 
give the most essential
contribution to this quantity, while the vacuum quarks are negligible.

So, the ratio $\Delta c /\Delta \Sigma $ can be easily calculated in 
the same line
as for ${f_{\eta '}^{(c)}}/{f_{\eta '}^{(u)}} $ \re{ratio} which leads to
\be 
\Delta c /\Delta \Sigma \,=\,-\frac{12}{5\rho^{2}}
\frac{1}{N_{f}6m_{c}^{2}} \,\sim\, -0.033.
\lab{ratio2}
\ee
With the use of \re{DS} it  means 
\be
\Delta c \,=\,-\, (0.01\,\sim\,0.016).
\lab{Dc}
\ee
Again the quantity $\Delta c$ in \re{Dc} is  given
at the scale of the size of the instanton.
On the scale $\approx m_c$
the account of the   anomalous dimension of the 
$g^3G\tilde{G}G$ operator  \ci{M84} leads to the same correction 
as in \re{correction}
and we have  
\be
\label{correctionDc}
\Delta c  (\mu \simeq m_c)
\simeq 1.5\Delta c (\mu \simeq \rho^{-1})\,=\,-\, (0.015\,\sim\,0.024).
\ee

\section{Conclusion} 

The problem of the virtual charm axial current contribution has been reduced to
the calculations of the specific gluon operator matrix elements  by the
application of the operator Schwinger method, developed for  QCD by the
ITEP group. Due to the
specific structure of this operator these matrix elements 
receive the main contribution from the instanton  background. 
The DP effective action approach, based on instanton model of QCD vacuum,
has been rederived and applied to the calculations of such types of  matrix
elements in chiral
limit, since the reliable answer may be obtained only in this limit. 

We have calculated  the coupling of
$\eta '$ with virtual charm axial current $f_{\eta '}^{(c)}$ \re{fc-4}.
The obtained value
 may provide a  satisfactory
 explanation of the experimental data
on the branching ratio for 
$B^\pm \rightarrow \eta^{\prime}  K^\pm$-decay as was shown recently by 
Cheng and Tseng
 \ci{CT98}. They demonstrated an impressive good explanation of these data 
 using  our value \re{fc-4} for $f_{\eta '}^{(c)}$.
 
This approach has been applied also to  the problem of the charm content 
of the nucleon spin $\Delta c.$ We conclude that 
$\Delta c$ \re{correctionDc} is one order of magnitude smaller than the 
analogous contribution of the
strangeness $\Delta s\,=\,-\, 0.11\,\pm\,0.03$ \ci{EK95}.
\\ \\
 We acknowledge the support of the COE program and a partial support by 
 the grant INTAS-96-0597, which enables M.M. to stay at RCNP of Osaka
 University and to perform the present study.

\clearpage
\def\theequation{\thesection.\arabic{equation}}
\def\thesection{A}
\setcounter{equation}{0}
\begin{center}{\bf APPENDIX}\end{center}

We calculate here the determinant of the Dirac operator 
which appears from the Gaussian integral over c-quarks.
This calculation was performed by using
 of the expansion of the effective action 
in the series of $G/m_c^2$ by Vainshtein et al. 
in Ref. \cite{VZNS83}.
However, we show here that careful calculations give a
different result from that of Vainshtein et al \cite{VZNS83}.
Throughout this appendix, 
we use the Euclidean convention
and consider up to the third order of $G$ in the expansion.
The contribution of the source $j_{\mu}$
which satisfies the field equation $D_{\nu}G_{\nu\mu}=j_{\mu}$
may be negligible,
as being consistent to the main part of this paper.

As explained in the calculation of 
$\langle 2m_c c^\dagger \gamma_5 c\rangle$,
the determinant of the Dirac operator, $\det\|\Dirac\|$,
must be regularized by the regulator mass $M$,
\begin{equation}
{\cal D}\equiv \det\left\|\frac{(\Dirac)(\fs{p}+iM)}
{(\fs{p}+im_c)(\Fs{\cP}+iM)}\right\|\;.
\end{equation}
Instead of the direct calculation of ${\cal D}$, 
it is more convenient to calculate the logarithm of ${\cal D}$.
The relation, $\ln\det\|A\|=\Tr\ln\|A\|$, 
is satisfied formally with the infinite-dimensional matrix $A$,
\begin{equation}
\ln{\cal D}
=\Tr\ln\left\|\frac{(\Dirac)(\fs{p}+iM)}
{(\fs{p}+im_c)(\Fs{\cP}+iM)}\right\|\;.
\end{equation}
Note that the symbol $\Tr$ in this appendix contains 
the trace over Lorentz and color indices and
the integral over the coordinate, i. e.
\begin{equation}
\Tr(\cdots)\equiv\tr_{L+C}\int d^{4}x\bra{x}\cdots\ket{x}\;,
\end{equation}
while in the main part of this paper it has been written as
$\Tr\equiv\tr_{L+C}$ only.

Let us consider the derivative of $\ln {\cal D}$ 
with respect to the c-quark mass $m_{c}$,
\begin{eqnarray}
\frac{1}{m_c}\frac{d}{dm_c}\ln{\cal D}
&=&\frac{1}{m_c}\Tr\left(\frac{i}{\Dirac}
-\frac{i}{\fs{p}+im_c}\right)
=\Tr\left(\frac{1}{\Klein+\frac{g}{2}\sG}
-\frac{1}{p^{2}+m_c^{2}}\right)\nonumber\\
&=&\Tr\left(\frac{1}{\Klein}-\frac{1}{p^{2}+m_c^{2}}\right)\nonumber\\
&&+\Tr\frac{1}{\Klein}\frac{g}{2}\sG\frac{1}{\Klein}
\frac{g}{2}\sG\frac{1}{\Klein}\\
&&-\Tr\frac{1}{\Klein}\frac{g}{2}\sG\frac{1}{\Klein}
\frac{g}{2}\sG\frac{1}{\Klein}\frac{g}{2}\sG\frac{1}{\Klein}+\cdots
\nonumber\\
&\equiv&I_0+I_2+I_3+\cdots\;,\nonumber
\end{eqnarray}
where the single $G_{\mu\nu}$ term vanishes 
because of the trace of the single $\sigma_{\mu\nu}$.

Calculations of both $I_{2}$ and $I_{3}$ are performed 
as we have done in the calculation of 
$\langle 2m_{c}c^{\dagger}\gamma_{5}c\rangle$,
except for $\gamma_{5}$.
Since we need up to ${\cal O}(G^{3})$ in our purpose,
each of them is calculated as follows:
\begin{eqnarray}
I_{3}(m_{c}^{2})
&=&-\Tr\frac{1}{\Klein}\frac{g}{2}\sG\frac{1}{\Klein}
\frac{g}{2}\sG\frac{1}{\Klein}\frac{g}{2}\sG\frac{1}{\Klein}\nonumber\\
&=&-\Tr\frac{1}{(\Klein)^{4}}\left(\frac{g}{2}\sG\right)^{3}\nonumber\\
&=&-\int d^{4}x\bra{x}\frac{1}{(\Klein)^{4}}\ket{x}
\,\tr_{L+C}\left\{\frac{g}{2}\sG(x)\right\}^{3}\nonumber\\
&=&-\frac{g^{3}}{2^{5}\cdot 3\pi^{2}m_c^{4}}\int d^{4}x
f_{abc}G^{a}G^{b}G^{c}(x)\;,
\end{eqnarray}
and
\begin{eqnarray}
I_{2}(m_{c}^{2})
&=&\Tr\frac{1}{\Klein}\frac{g}{2}\sG\frac{1}{\Klein}
\frac{g}{2}\sG\frac{1}{\Klein}\nonumber\\
&=&\frac{g^{2}}{4}\int d^{4}x\left\{
\bra{x}\frac{1}{(\Klein)^{3}}\ket{x}\tr_{L+C}\left(\sG\right)^{2}
+\bra{x}\frac{1}{(\Klein)^{4}}\ket{x}\tr_{L+C}\sG\cdot D^{2}\sG\right.
\nonumber\\
&&\;\;\;\;\;\;\;\;\;\;\;\;\;\;\;\;\;
\left.-4\bra{x}\frac{1}{(\Klein)^{5}}\cP_{\nu}\cP_{\mu}\ket{x}
\tr_{L+C}\sG\cdot D_{\nu}D_{\mu}\sG\right\}\nonumber\\
&=&\frac{g^{2}}{4}\int d^{4}x\left\{
\frac{1}{2^{5}\pi^{2}m_c^{2}}\tr_{L+C}\left(\sG\right)^{2}
+\frac{1}{2^{6}\cdot 3\pi^{2}m_c^{4}}
\tr_{L+C}\sG\cdot D^{2}\sG\right\}\nonumber\\
&=&\frac{g^{2}}{2^{5}\pi^{2}m_c^{2}}\int d^{4}x G^{a}G^{a}(x)
+\frac{g^{3}}{2^{6}\cdot 3\pi^{2}m_c^{4}}
\int d^{4}x f_{abc}G^{a}G^{b}G^{c}(x)
\;.
\end{eqnarray}
Here we have used the following relations,
\begin{eqnarray}
&\tr_{L+C}\left(\sG\right)^{3}
&=-2^{5}i\cdot\tr_{C}GGG
=2^{3}f_{abc}G^{a}G^{b}G^{c}\nonumber\\
&\tr_{L+C}\left(\sG\right)^{2}&=2^{3}tr_{C}GG=2^{2}G^{a}G^{a}\\
&D^{2}\sG&=2ig\sigma_{\mu\nu}G_{\mu\alpha}G_{\alpha\nu}\nonumber
\end{eqnarray}
under neglecting the source $j_{\mu}$
which satisfies $D_{\nu}G_{\nu\mu}=j_{\mu}$.

The calculation of $I_{0}$ is rather technical
since it needs careful treatments of
traces of infinite-dimensional matrices.
First, we consider the translational invariance of $I_{0}$.
It means that 
$I_{0}$ does not change under a shift of the operator $\cP_{\mu}$ 
by an arbitrary vector $q_{\mu}$, 
\begin{eqnarray}
I_{0}(m_{c}^{2})
&=&\Tr\left(\frac{1}{\Klein}
-\frac{1}{p^{2}+m_{c}^{2}}\right)\nonumber\\
&=&\Tr\; e^{iqX}\left(\frac{1}{\Klein}
-\frac{1}{p^{2}+m_{c}^{2}}\right)e^{-iqX}\label{eq:I0}\\
&=&\Tr\left\{\frac{1}{(\cP-q)^{2}+m_{c}^{2}}
-\frac{1}{(p-q)^{2}+m_{c}^{2}}\right\}\;.\nonumber
\end{eqnarray}
Here $e^{iqX}$ is the shift operator in momentum space,
and $X_{\mu}$ is the coordinate operator
which satisfies the algebra (\ref{eq:op1}).
Now, we expand the right hand side of (\ref{eq:I0})
into the series in $q_{\mu}$.
The momentum shift of $(\Klein)^{-1}$ is expanded as
\begin{eqnarray}
\lefteqn{e^{iqX}\frac{1}{\Klein}\,e^{-iqX}}\nonumber\\
&=&\sum_{n=0}^{\infty}\,\frac{1}{n!}\,\underbrace{
[\,iqX,[\,iqX,\cdots\,[\,iqX\,,\,\frac{1}{\Klein}\,]\cdots\,]\,]
}_{iqX\;\;{\rm appears}\;\;n\;\;{\rm times}}\nonumber\\
&=&\frac{1}{\Klein}+[\,iqX\,,\,\frac{1}{\Klein}\,]
+\frac{1}{2}[\,iqX,[\,iqX\,,\,\frac{1}{\Klein}\,]\,]\,+\cdots\;.
\end{eqnarray}
When one apply the trace $\Tr$ with both sides of this equation,
the first term of the right hand side 
is equal to the left hand side
and both terms do not depend at all on the vector $q_{\mu}$.
This means that other terms 
which depend on any power of $q$ 
are identically equal to zero, i. e.
\begin{equation}
\Tr\,[\,iqX\,,\,\frac{1}{\Klein}\,]
=\Tr\,[\,iqX,[\,iqX\,,\,\frac{1}{\Klein}\,]\,]
=\cdots
=0\;\;\;\;,\;\;\;\;{\rm for\;any}\;q_\mu\;.
\end{equation}
Consider the coefficient of the term
which is the second order of $q_{\mu}$:
\begin{eqnarray}
\lefteqn{\Tr\,[\,iqX,[\,iqX\,,\,\frac{1}{\Klein}
\,-\,\frac{1}{p^2+m_c^2}\,]\,]}\nonumber\\
&=&-2q_\mu q_\nu
\left[\,\delta_{\mu\nu}\Tr\left\{\frac{1}{(\Klein)^2}
\,-\,\frac{1}{(p^2+m_c^2)^2}\right\}\right.\nonumber\\
&&\;\;\;\;\;\;\;\;\;\;\;\;\;\;
\left.-4\Tr\left\{
\frac{1}{(\Klein)^2}\cP_\mu\frac{1}{\Klein}\cP_\nu
\;-\;\frac{p_\mu p_\nu}{(p^2+m_c^2)^3}
\right\}\right]\;.
\end{eqnarray}
Since $q_{\mu}$ is an arbitrary vector, we may choose 
the average over the direction of $q_{\mu}$, i. e. 
$q_{\mu}q_{\nu}=\frac{1}{4}q^{2}\delta_{\mu\nu}$.
As explained already, the coefficient of $q^{2}$ is equal to zero.
The corresponding equation looks as
\begin{equation}
\Tr\left\{\frac{1}{(\Klein)^2}\,-\,\frac{1}{(p^2+m_c^2)^2}\right\}
=\Tr\left\{\frac{1}{(\Klein)^2}\cP_\mu\frac{1}{\Klein}\cP_\mu
\;-\;\frac{p^2}{(p^2+m_c^2)^3}\right\}.\label{eq:klein2}
\end{equation}
On the other hand, using the relation $(\Klein)(\Klein)^{-1}=1$,
\begin{eqnarray}
\lefteqn{\Tr\left\{\frac{1}{(\Klein)^2}\,
-\,\frac{1}{(p^2+m_c^2)^2}\right\}}\nonumber\\
&=&\Tr\left\{\frac{1}{(\Klein)^3}\cP^{2}\,
-\,\frac{p^{2}}{(p^2+m_c^2)^3}\right\}
\,+\,m_{c}^{2}\Tr\left\{\frac{1}{(\Klein)^3}\,
-\,\frac{1}{(p^2+m_c^2)^3}\right\}\;.
\end{eqnarray}
Hence, substituting it to Eq. (\ref{eq:klein2}), 
we obtain the relation
\begin{eqnarray}
\lefteqn{\Tr\left\{\frac{1}{(\Klein)^3}\,
-\,\frac{1}{(p^2+m_c^2)^3}\right\}}\nonumber\\
&=&\frac{1}{m_{c}^{2}}
\Tr\frac{1}{(\Klein)^2}[\,\cP_{\mu}\,,
\,\frac{1}{(\Klein)}\,]\,\cP_{\mu}\nonumber\\
&=&\frac{1}{m_{c}^{2}}
\Tr\sum_{n=1}^{\infty}\frac{(-1)^{n}}{(\Klein)^{n+3}}
\cP_{\mu}\,\underbrace{[\,\cP^{2}\,,\,
[\,\cP^{2}\,,\cdots\,
[\,\cP^{2}\,,\cP_{\mu}\,]\,
\cdots\,]\,]}_{\cP^{2}\;\;{\rm appears}\;\;n\;\;{\rm times}}
\nonumber\\
&=&\frac{1}{m_{c}^{2}}\Tr\frac{1}{(\Klein)^4}
\left\{-\cP_{\mu}[\,\cP^{2}\,,\cP_{\mu}\,]
\,-\frac{1}{\Klein}[\,\cP^{2}\,,\cP_{\mu}\,]\,
[\,\cP^{2}\,,\cP_{\mu}\,]\right.\nonumber\\
&&\;\;\;\;\;\;\left.
+\frac{1}{(\Klein)^2}[\,\cP^{2}\,,\cP_{\mu}\,]\,
[\,\cP^{2}\,,\,[\,\cP^{2}\,,\cP_{\mu}\,]\,]\right.\\
&&\;\;\;\;\;\;\left.
+\frac{1}{(\Klein)^3}[\,\cP^{2}\,,\,[\,\cP^{2}\,,\cP_{\mu}\,]\,]\,
[\,\cP^{2}\,,\,[\,\cP^{2}\,,\cP_{\mu}\,]\,]\,\cdots\,\right\}\;.
\nonumber
\end{eqnarray}
To integrate the right hand side 
over momentum space up to the desired order of $G$,
we need only two commutation relations
under $j_{\mu}$ being neglected,
\begin{equation}
\begin{array}{cl}
[\,\cP^{2}\,,\cP_{\mu}\,]&=2ig\cP_{\nu}G_{\nu\mu}\\
\ [\,\cP^{2}\,,\,[\,\cP^{2}\,,\,\cP_{\mu}\,]\,]
&=-4g\cP_{\alpha}\cP_{\nu}D_{\alpha}G_{\nu\mu}
-4g^{2}\cP_{\alpha}G_{\alpha\nu}G_{\nu\mu}
+2ig\cP_{\alpha}D^{2}G_{\alpha\mu},\\
\end{array}
\label{eq:comP2}
\end{equation}
since the product of $\cP_{\mu}$ and the higher commutators 
which could give the ${\cal O}(G^{3})$ terms
can be replaced by the product of these commutators (\ref{eq:comP2})
with a suitable sign in the trace $\Tr$.
Thus,
\begin{eqnarray}
\lefteqn{\Tr\left\{\frac{1}{(\Klein)^3}\,
-\,\frac{1}{(p^2+m_c^2)^3}\right\}}\nonumber\\
&=&
-\frac{g^{2}}{2^{5}\cdot 3\pi^{2}m_{c}^{6}}\int d^{4}x G^{a}G^{a}(x)
-\frac{g^{3}}{2^{3}\cdot 3\cdot 5\pi^{2}m_{c}^{8}}
\int d^{4}x f_{abc}G^{a}G^{b}G^{c}(x)\;.
\end{eqnarray}
The integration of this result over squared masses twice gives
\begin{eqnarray}
I_{0}(m_{c}^{2})
&=&2\int_{m_{c}^{2}}^{\infty}dm_{1}^{2}
\int_{m_{1}^{2}}^{\infty}dm_{2}^{2}
\Tr\left\{\frac{1}{(\cP^{2}+m_{2}^{2})^3}\,
-\,\frac{1}{(p^2+m_2^2)^3}\right\}\nonumber\\
&=&
-\frac{g^{2}}{2^{5}\cdot 3\pi^{2}m_{c}^{2}}\int d^{4}x G^{a}G^{a}(x)
-\frac{g^{3}}{2^{3}\cdot 3^{2}\cdot 5\pi^{2}m_{c}^{4}}
\int d^{4}x f_{abc}G^{a}G^{b}G^{c}(x)\;.
\end{eqnarray}
Finally, we obtain the determinant
\begin{eqnarray}
{\cal D}
&=&\exp\frac{1}{2}\int_{m_{c}^{2}}^{M^{2}}dm^{2}\,
\left\{I_{0}(m_{c}^{2})+I_{2}(m_{c}^{2})+I_{3}(m_{c}^{2})\right\}
\nonumber\\
&=&\exp\int d^{4}x\left\{
\frac{g^{2}}{2^{5}\cdot 3\pi^{2}}\ln\frac{M^{2}}{m_{c}^{2}}G^{a}G^{a}(x)
-\frac{23g^{3}}{2^{7}\cdot 3^{2}\cdot 5\pi^{2}m_{c}^{2}}
f_{abc}G^{a}G^{b}G^{c}(x)
\right\}\;.
\end{eqnarray}

To compare with the result of Vainshtein et al. \cite{VZNS83},
we needs to obtain the effective action written in the Minkowski metric.
In our case, the Euclidean effective action 
is defined as
\begin{equation}
S_{eff}^{E}=\ln{\cal D}.
\end{equation}
According to the previous convention
for the transition between the Minkowskian and the Euclidean space,
\begin{equation}
(GG)_{E}=(GG)_{M}\;\;,\;\;(GGG)_{E}=-(GGG)_{M}\;.
\end{equation}
Then the Minkowskian effective action is
\begin{equation}
S_{eff}^{M}=\int d^{4}x 
\left\{
-\frac{g^{2}}{2^{4}\cdot 3\pi^{2}}\ln\frac{M^{2}}{m_{c}^{2}}\tr_{C}GG(x)
+\frac{23ig^{3}}{2^{5}\cdot 3^{2}\cdot 5\pi^{2}m_{c}^{2}}\tr_{C}GGG(x)
\right\}\;,
\end{equation}
Here, $\tr_{C}$ is explicitly written
in order to compare with the convention of Vainshtein et al. \cite{VZNS83}.
The coefficient of the second term 
is different from their result \cite{VZNS83} with factor $23/2$.


\begin{thebibliography}{9}

 \bibitem{CLEO1} J. Smith (CLEO Collaboration), talk presented at the
1997 ASPEN winter conference on Particle Physics, Aspen, Colorado,
1997; 
S. Anderson et al.~(CLEO Collaboration), CLEO CONF 
97-22a and EPS 97-333 (1997).
                       
\bibitem{CLEO2}
B.H. Behrens et al. (CLEO Collaboration), CLNS 97/1536,
CLEO 97-31, hep-ex/9801012 (1998),
 {\it Phys.Rev.Lett.}{\bf 80},3710-3714,1998.

\bibitem{HZ}I. Halperin and A. Zhitnitsky, hep-ph/9704412,
 {\it Phys.Rev.}{\bf D56},7247 (1997).

\bibitem{SZ}E. V. Shuryak, A. Zhitnitsky, hep-ph/9706316, 
     {\it Phys.Rev.}{\bf D57},2001(1998).

\bibitem{AG97}
A. Ali and C. Greub, hep-ph/9707251, {\it Phys. Rev.} {\bf D57}, 2996 (1998).

\bibitem{ACGK97}A. Ali, J. Chay, C. Greub and P. Ko, hep-ph/9712372,
{\it Phys. Lett.} {\bf B424}, 161 (1998).

\bibitem{FK97}T. Feldmann and P. Kroll, preprint WUB 97-28, hep-ph/9711231.

\bibitem{FK98}
T. Feldmann, P. Kroll and B. Stech, preprint WUB 98-2, HD-THEP-98-5,
 hep-ph/9802409.

\bibitem{F}
H. Fritzsch, preprint LMU 08/97, CERN-TH-97-200,
 hep-ph/9708348, {\it Phys. Lett.} {\bf B415}, 83 (1997).

\bibitem{CT}
Hai-Yang Cheng, B. Tseng, preprint IP-ASTP 03-97, NTU-TH-97-08,
 hep-ph/9707316, {\it Phys.Lett.}{\bf B415},263-272,1997.

\bibitem{CT98}
Hai-Yang Cheng, B. Tseng, preprint IP-ASTP 01-98, 
 hep-ph/9803457.

\bibitem{DKY}
Dongsheng Du, C.S.Kim, Yadong Yang, preprint BIHEP-Th/97-15, SNUTP 97-150, YUMS
97-029, hep-ph/9711428, {\it Phys. Lett.} {\bf B426}, 133 (1998).

\bibitem{P}
A.A.Petrov, preprint JHU-TIPAC-97016, UMHEP-446,
hep-ph/9712497, {\it  Phys.Rev.}{\bf D58},054004,1998.

\bibitem{K98} 
Alex Kogan, hep-ph/9806266.

\bibitem{EK95}
J. Ellis, M. Karliner, {\it Phys. Lett.} {\bf B341}, 397 (1995),
 CERN-TH/95-279, TAUP-2297-95, hep-ph/9510402.

\bibitem{Mulders98}
P.J. Mulders, hep-ph/9806314.

\bibitem{Ioffe98}
B. L. Ioffe, hep-ph/9804238.

\bibitem{HZ97-dis}
I. Halperin, A. Zhitnitsky, hep-ph/9706251.  

\bibitem{BS97}
A. Blotz, E. Shuryak, hep-ph/9710544.

\bibitem{Schw} J. Schwinger, {\it Phys. Rev.} {\bf 82}, 664 (1951). 

\bibitem{Adl} S. Adler, {\it Phys. Rev.} {\bf 177}, 2426 (1969). 

\bibitem{BJa} J. S. Bell and R. Jackiw, {\it Nuovo Cim.} {\bf 60, ser. A}, 47 
(1969).

\bibitem{Fujikawa} K. Fujikawa,
    {\it  Phys. Rev.} {\bf D 21}, 2848 (1980) \\
    in Proc. NATO Advanced Research Workshop "Super Field Theories",
    July 25-30, 1986.

\bibitem{FMP} B. Faizullaev, M. M. Musakhanov, N. K. Pak, 
 {\it  Phys.   Lett.} {\bf  B 361}, 155 (1995). 

\bibitem{Shi} M. A. Shifman,  {\it Sov. Phys. Usp. } {\bf 32}, 289 (1989);
{\it Phys. Rep.} {\bf 209}, 341 (1991).  

\bibitem{MK} M. M. Musakhanov, F. C. Khanna,  
{\it  Phys.  Lett.} {\bf  B 395}, 298 (1997).
                        
\bibitem{SM} E.Di Salvo, M.M. Musakhanov, hep-ph/9706537, 
 {\it Eur. Phys. J. }{\bf C}(Online publication: March 10,1998).

\bibitem{DP} D. Diakonov and V. Petrov, "Spontaneous breaking of 
chiral symmetry in the instanton vacuum", LNPI preprint LNPI-1153(1986);
{\it in} "Hadron matter under extreme conditions", (Kiev),1986, p. 192.

\bibitem{DPW} D. I. Diakonov,  M. V. Polyakov, C. Weiss, 
{\it  Nucl. Phys. } {\bf  B 461}, 539 (1996). 

\bibitem{LB79} C. Lee, W. A. Bardeen, 
{\it  Nucl. Phys. } {\bf  B 153}, 210 (1979). 

\bibitem{Tokarev}  V. F. Tokarev, Instantons and Colour Screening,
preprint INP P-0406, Moscow 1985;
{\it Soviet J. Theor. Math. Phys.}, {\bf  73}, 223 (1987). 

\bibitem{Shu82}E. V. Shuryak,  
{\it  Nucl. Phys. } {\bf  B 203}, 93, 116 (1982). 

\bibitem{DP84} D. Diakonov and V. Petrov,  
{\it  Nucl. Phys. } {\bf  B 245}, 259 (1984). 

\bibitem{VZNS83} 
A. I. Vainshtein, V. I. Zakharov, V. A. Novikov, M. A. Shifman,
                Sov.J.Nucl.Phys. {\bf 39}, 77 (1984).

\bibitem{M84} A. Yu. Morozov, 
              Sov.J.Nucl.Phys. {\bf 40}, 505 (1984).

\bibitem{WY90} M. Wakamatsu, H. Yoshiki,
{\it  Nucl. Phys. } {\bf  A524}, 561 (1991).

\bibitem{BPG93}
A. Blotz, M. Polyakov, K. Goeke, 
{\it  Phys.  Lett.} {\bf  B 302}, 151 (1993). 

\bibitem{D98} D. Diakonov, Chiral Quark-Soliton Model, hep-ph/9802298. 

\end{thebibliography}
\end{document}